\pdfoutput=1 

\documentclass[]{spie}  
\usepackage[]{graphicx}

\title{MUSTANG: 90 GHz Science with the Green Bank Telescope.} 

\author{S.R.~Dicker\supit{a},
P.M.~Korngut\supit{a},
B.S.~Mason\supit{b},
P.A.R.~Ade\supit{e},
J.~Aguirre\supit{a},
T.J.~Ames\supit{d},\\
D.J.~Benford\supit{d},
T.C.~Chen\supit{d},
J.A.~Chervenak\supit{d},
W.D.~Cotton\supit{b},
M.J.~Devlin\supit{a},\\
E.~Figueroa-Feliciano\supit{f},
K.D.~Irwin\supit{g},
S.~Maher\supit{d},
M.~Mello\supit{c},
S.H.~Moseley\supit{d},\\
D.J.~Tally\supit{d},
C.~Tucker\supit{e},
and S.D.~White\supit{c}
\skiplinehalf
\supit{a}University of Pennsylvania, 209 S. 33rd St.,  Philadelphia, PA 19104, USA;\\
\supit{b}National Radio Astronomy Observatory, 520 Edgemont Rd, Charlottsville, VA 22903, USA;\\
\supit{c}National Radio Astronomy Observatory, Green Bank, WV 24944, USA;\\
\supit{d}NASA Goddard Space Flight Center, Greenbelt, MD 20771, USA;\\
\supit{e}School of Physics and Astronomy, Cardiff University, 5 The Parade, Cardiff, CF24 3AA, UK;\\
\supit{f}Department of Physics, Massachusetts Institute of Technology, Cambridge, MA 02139, USA;\\
\supit{g}National Institute of Standards and Technology, 325 Broadway, Boulder, CO 80303, USA}

\begin{document} 
  \maketitle 
\newcommand{\apj}{ApJ}
\newcommand{\aj}{AJ}
\newcommand{\ao}{Applied Optics}
\newcommand{\apjl}{ApJ}
\newcommand{\mnras}{MNRAS}
\newcommand{\aap}{A\&A}
\begin{abstract}
MUSTANG is a 90~GHz bolometer camera built for use as a facility
instrument on the 100~m Robert C. Byrd Green Bank radio
telescope (GBT)\@. MUSTANG has an 8 by 8 focal plane array of transition edge
sensor bolometers read out using time-domain multiplexed SQUID electronics.
As a continuum instrument on a large single dish MUSTANG has a
combination of high resolution ($8''$) and good sensitivity to
extended emission which make it very competitive for a wide range of
galactic and extragalactic science.  Commissioning finished in January
2008 and some of the first science data have been collected.  

\end{abstract}



\section{INTRODUCTION}
\label{sec:intro}
Advances in radio astronomy have been driven by the development
of more sensitive receivers.  However at higher
frequencies  ($>$60~GHz) the 
receiver noise is beginning to be dominated by the random arrival of
photons so  building lower noise detectors will no longer produce a
more sensitive receiver.  When this happens the only way to improve on 
an instrument is to put in more detectors.  Large arrays of detectors
 also have many other advantages over receivers with just a few pixels.  
For example many systematic effects such as 
emission from the atmosphere will be common mode to all detectors and
so can be  removed in data analysis, reducing the need for chopping or
rapid scanning, techniques  that lower observing efficiency.  

The Robert C. Byrd 
Green Bank Telescope (GBT) is a 100~m diameter radio telescope
located in West Virginia, USA\@.  It has an off-axis
Gregorian design and an active primary mirror consisting of 2004 panels.
A semi-empirical model of how the telescope deforms under gravity 
keeps the shape of the mirror
accurate to 390~$\mu $m RMS with an eventual target of 240~$\mu $m.  
At 90~GHz, a 240~$\mu $m RMS corresponds to a surface efficiency 
of 43\% (or an 
effective surface area of 2700~m$^2$).  This large surface area
combined with a 90~GHz resolution of 8$''$ FWHM makes the GBT a unique
instrument for a wide range of science. Although interferometers 
can achieve higher resolution  they have little sensitivity 
to extended emission and their total collecting area is often small.

MUSTANG is a bolometer camera designed for the GBT\@. It operates at
the GBT's Gregorian focus where a rotating turret houses seven other
receivers.  This allows users to change between receivers in just 
a few minutes.  Using the GBT at 90~GHz requires
benign weather with low wind, low opacity and observations must be at
night so uneven heating from the Sun does not distort the dish.
  Currently the GBT operates on a flexible scheduling scheme
with back-up low frequency projects getting the use of the telescope
when high frequency observations are not possible.  True dynamic
scheduling will be implemented shortly.  During the winter of 2008 the
MUSTANG team found that approximately half of available nights had
reasonable weather for 90~GHz observing. 

\subsection{Science Goals}
MUSTANG's current sensitivity allows it to map a $3' \times 3'$ patch 
of sky to 2.5~mJy/beam in one hour.  We expect this to improve by a
factor of 10 over the next few years. 
 As a user instrument these sensitivities will allow for a
wide range of new galactic, extragalactic, and solar system science.  
Some examples where MUSTANG will make an impact are discussed next.

\subsubsection{Galaxy clusters and the Sunyaev-Zel'dovich Effect}
In the next few years experiments such as ACT\cite{ACT} and the
SPT\cite{SPT} will finish blind surveys for clusters 
using the Sunyaev-Zel'dovich Effect (SZE).  These surveys aim to
constrain the evolution of dark energy by measuring the abundance of
clusters as a function of redshift.  In interpreting these surveys it
is important to understand the relation of mass to SZE observables and
to constrain the physics contributing to the scatter in these
relations.  For instance, high angular
resolution X-ray observations have revealed relic ``cold
fronts'' in the intra-cluster plasma\cite{Markevitch2000} and bubbles
inflated by AGN ejecta.\cite{Birzan} Spatially resolved SZE
measurements are essential to understanding these phenomena and their
effect on the lower-resolution SZE measurements from the surveys.

\subsubsection{Star \& planet formation}
The details of how opaque clouds of dust and gas collapse into stars
have been the subject of a number of 
studies.\cite{Weintraub1989}\,\cite{Beckwith1990}\,\cite{Mannings1994}
There are many questions about what triggers the collapse of a cloud,
what determines the mass of the stars formed and the physics of accretion
disks known to form around most protostars.  When combined with higher
frequency data, accurate photometry at 90~GHz
will constrain the spectral index of the dust emission from any
disks observed.  From this, the size of dust grains can be found,
restricting some theories of planet formation.

\section{INSTRUMENT DESCRIPTION}
At the heart of MUSTANG is an eight by eight array of Transition Edge
Sensor (TES) bolometers read out using time-domain SQUID
multiplexing electronics. The pixels are used without feedhorns and
are spaced by approximately half a beam width  ($4''$)
on the sky to give
a $32''$\ square field of view.  This arrangement ensures that the sky is
fully sampled in a single pointing of the GBT.  To avoid artifacts in
maps,  multiple observations of different points on the sky should be made
on time-scales less than typical drifts in the sky brightness 
(every few seconds).   With a fully sampled array, the scan speeds
and accelerations required to do this are reduced.  On a structure
the size of the GBT this is a great advantage.

TES bolometers use a voltage biased superconducting film to measure
optical power.  Electrical power heats the TES up above the bath
temperature  to its transition temperature, $\rm T_c$,  
where a natural feedback mechanism keeps the temperature constant.
An increase in the optical power being absorbed by the bolometer
warms the TES and its resistance increases. Consequently, the current flowing
through the device drops reducing the amount of electrical heating so
that the sum of the electrical and optical power (and hence the
temperature of the device) remains almost constant. 
Conversely, a decrease in optical power results in a decrease in
resistance and an increase in electrical heating.
 By measuring the current flowing through the TES you
measure the optical loading on the bolometer.  The strong
electrothermal feedback means that a TES bolometer has a response time
faster than its thermal time constant, its heat capacity divided by
the thermal conductivity from the bolometer to the cold sink.  The
phonon noise in a regular bolometer increases as the square root of
the thermal conductivity and the temperature, so for a given set of
operating conditions and target noise levels, an optimally designed
TES bolometer will be faster than an optimally designed semiconductor
bolometer\cite{TES_review}.  To map large areas  MUSTANG must scan
at tens of beamwidths per second so the faster time constants
intrinsic to TES bolometers made them more attractive for use in MUSTANG.  Another
attractive feature of TES bolometers is that they are of low impedance
devices making them less susceptible to microphonics in the readout
electronics. 

  The TES film is made of two layers, a superconducting metal
(in our case Molybdenum) and a normal metal (in our case gold).  The
proximity effect lowers the transition temperature of the
superconductor.  The thickness of the layers can be varied to give a
wide range of possible transition temperatures.  The exact choice of 
$\rm T_c$ depends on the desired detector properties but is limited by the choice of refrigeration used in the receiver. MUSTANG's detectors are cooled to 300~mK using a helium-3 sorption refrigerator backed by a helium-4 sorption refrigerator.  The helium-4
 refrigerator is, in turn, backed by a two stage pulse tube. Details of
 the cryogenic design and performance are given below in
 section~\ref{sec:cryo}. 

\subsection{Detector Design}

   \begin{figure}
   \begin{center}
   \begin{tabular}{c}
   \includegraphics[height=10cm]{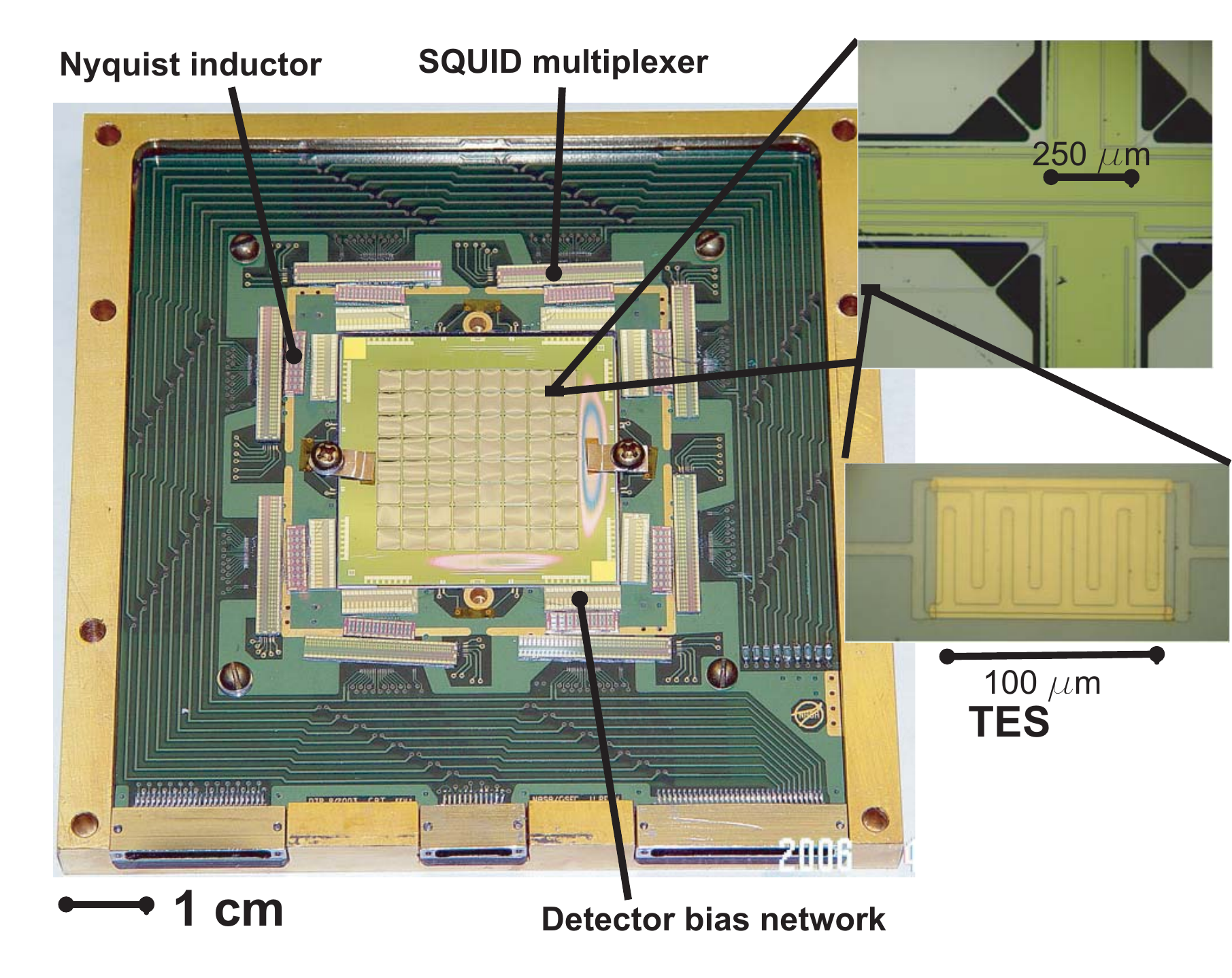}
   \end{tabular}
   \end{center}
   \caption[Array] 
   { \label{fig:array} MUSTANG's detector array and its multiplexed
     readout electronics.  The 35~mm square detector chip is mounted
     in a square hole at the center of a G10 circuit board.
     Superconducting aluminum wire
     bonds connect the detector chip to the multiplexing and bias chips that
     are glued to the G10 circuit board. The array is heat sunk to
     the circuit board
     using many gold bonds.  Only 9 out of 32 address
     lines are used and the usable part of the focal plane is much
     bigger than the current array so there is room to expand to more
     pixels. Shown in the top corner is an enlargement of the corners of
     four pixels, showing the 10$\mu $m wide legs that hold the
     membranes to the support structure on which the wiring can be
     seen. Below this is a close-up of the TES showing the gold bars that
     help suppress excess noise\cite{TES_noise}.}

   \end{figure} 

A photograph of the MUSTANG detector package is shown in
figure~\ref{fig:array}.  Each bolometer consists of a 1~$\mu $m thick,
2.88~mm square membrane. This is suspended by four 10~$\mu $m wide
legs from a frame 0.25~mm wide.  Pixels are spaced by 3.2mm.
 The array was micromachined out of a 450~$\mu$m 
thick silicon wafer at Goddard Space
Flight Center\cite{Benford2004}.    The pixels are coated with an
absorbing layer of bismuth which has an impedance matched to free
space (400$\Omega$).  The TES is 
deposited close to the edge of the membrane and
superconducting aluminum traces run out along two of the legs that
suspend the membrane and along the supporting beams to the edge of the
array chip.  When laying out the design care was taken so that only
traces from neighboring pixels run close to each other.  This reduces
the effect of electrical cross talk as the neighboring pixels observe
overlapping parts of the sky.

To maximize absorption efficiency there is a reflecting backshort
$\lambda /4$ behind the array.  It is constructed of gold plated
alumina which has a similar thermal expansion coefficient to the array so stress during
cool down is minimized.  The alumina is epoxied to flexible
copper mounts in the bottom of the copper box.

Two of the most important parameters that affect the noise in a TES
bolometer are $\rm T_c$ and $G$, the conduction from the membrane to
the bath.  These parameters also set the saturation power, a
loading above which the TES will not function.  Unfortunately values
for these parameters that give low noise also give low
saturation powers. Bolometer design is inevitably a trade off
between good performance in the best weather and the ability to
operate in a wider range of weather conditions.  For MUSTANG 
the expected loading
per detector was calculated to be between 2 and 9~pW while the
photon noise limit was calculated to
be 2.4--$6\times 10^{-17}\ \mbox{W}/\sqrt{\mbox{Hz}}$, depending on
weather and elevation.   So as to allow for
headroom, target saturation powers of 12~pW and a target NEPs
of $2\times 10^{-17}\ \mbox{W}/\sqrt{\mbox{Hz}}$ were chosen.  Later in the
design process it was found that pixels with a 12~pW saturation power
would be very fragile so the design saturation power was raised to
25~pW.  To give plenty of headroom for operating off the helium-3
refrigerators, the transition temperature was chosen to be 450~mK. 

To read out the TES we use the time-domain SQUID multiplexers made by
NIST\cite{mux_paper} and the Mark III electronics developed at
Goddard\cite{Forgione2004}.  Eight multiplexer chips are arranged around
the array along with resistor networks, used to provide a constant
voltage bias across each TES, and Nyquist inductors that limit high
frequency noise.  Each multiplexer chip and the electronics used to
drive them can read up to 32 detectors so future expansion of the
array is possible.  Currently, each chip reads out 8 bolometers and a
dark channel which is used to detect $1/f$ noise in the electronics.
To avoid producing interference in other GBT receivers, all room
temperature electronics are housed in RFI tight boxes.  Tests in an
anechoic chamber showed that, above 30~MHz, no signal larger than
-65~dBm was emitted. 
 
\subsection{Cryogenic Design and Performance}   
\label{sec:cryo}

\begin{figure}
   \begin{center}
   \begin{tabular}{c}
   \includegraphics[height=9cm]{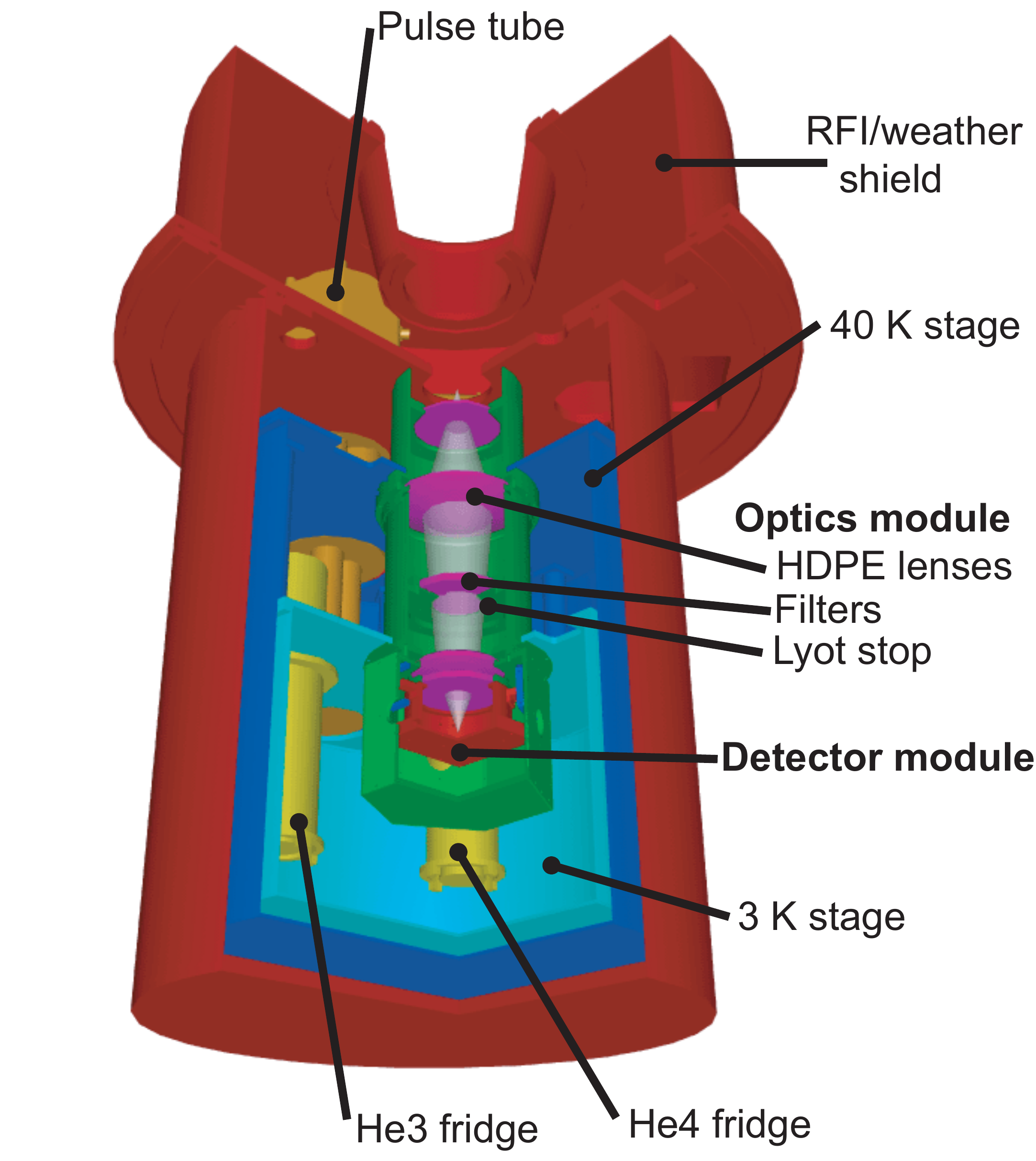}
   \end{tabular}
   \end{center}
   \caption[cryostat]{\label{fig:cryostat} A cross section of MUSTANG
     showing the arrangement of the optics and cryogenics.}
\end{figure}

 MUSTANG's detectors are cooled to below 300~mK using a helium-3
sorption refrigerator backed by a helium-4
sorption refrigerator (700~mK). 
This in turn, is backed up with a two-stage pulse tube that has temperature
stages at 3~K and 40~K\cite{cryo_paper}.  
  By using a pulse tube, expendable cryogens
are not needed, greatly simplifying the logistics of operating on
the GBT where access to receivers is limited.  This comes at the
expense of a dependence of operating temperature on elevation as the pulse
tube's performance drops off when it is tipped by more than
45$^\circ$. Consequently,
observations at elevations below 20$^\circ $\ are limited.  However, the
90~GHz opacity of the atmosphere at these elevations limits
observations so little is lost.  Pulse tubes have the
advantage of much lower vibrations than other types of cryocoolers,
(such Gifford-McMahon). 

In the lab, after a 3 hour cycling procedure, 
 the hold time for the helium-4 refrigerator is 48 hours
and the helium-3 refrigerator can stay cold for over 96 hours.  On the
telescope MUSTANG is constantly being tipped and cycling is often
carried out with the receiver at 45 degrees from the vertical.  In
order to ensure successful cycling of the sorption refrigerators in as
little time as possible, a second cycling procedure was developed so that the
cycle time was halved to 1.5 hours. This cycle was able to complete
with the GBT tipped over to elevations as low as 30 degrees. This
cycle procedure has a 
reduced the hold time for the helium-4 refrigerator of 14 hours,
however, this is more than enough for most observing sessions.  When
the helium-4 refrigerator runs out the temperature of the helium-3
stage increases by 15~mK and becomes more sensitive to drifts in the
temperature of the pulse tube.  The array still functions and data
 taken in this state are still of good quality.  All cycles are
 completed under the control of a computer
and can be started automatically when the helium-3 or helium-4
refrigerators run out or on demand.
   
\subsection{Optical Design}
\begin{figure}
   \begin{center}
   \begin{tabular}{c}
   \includegraphics[height=9cm,angle=0]{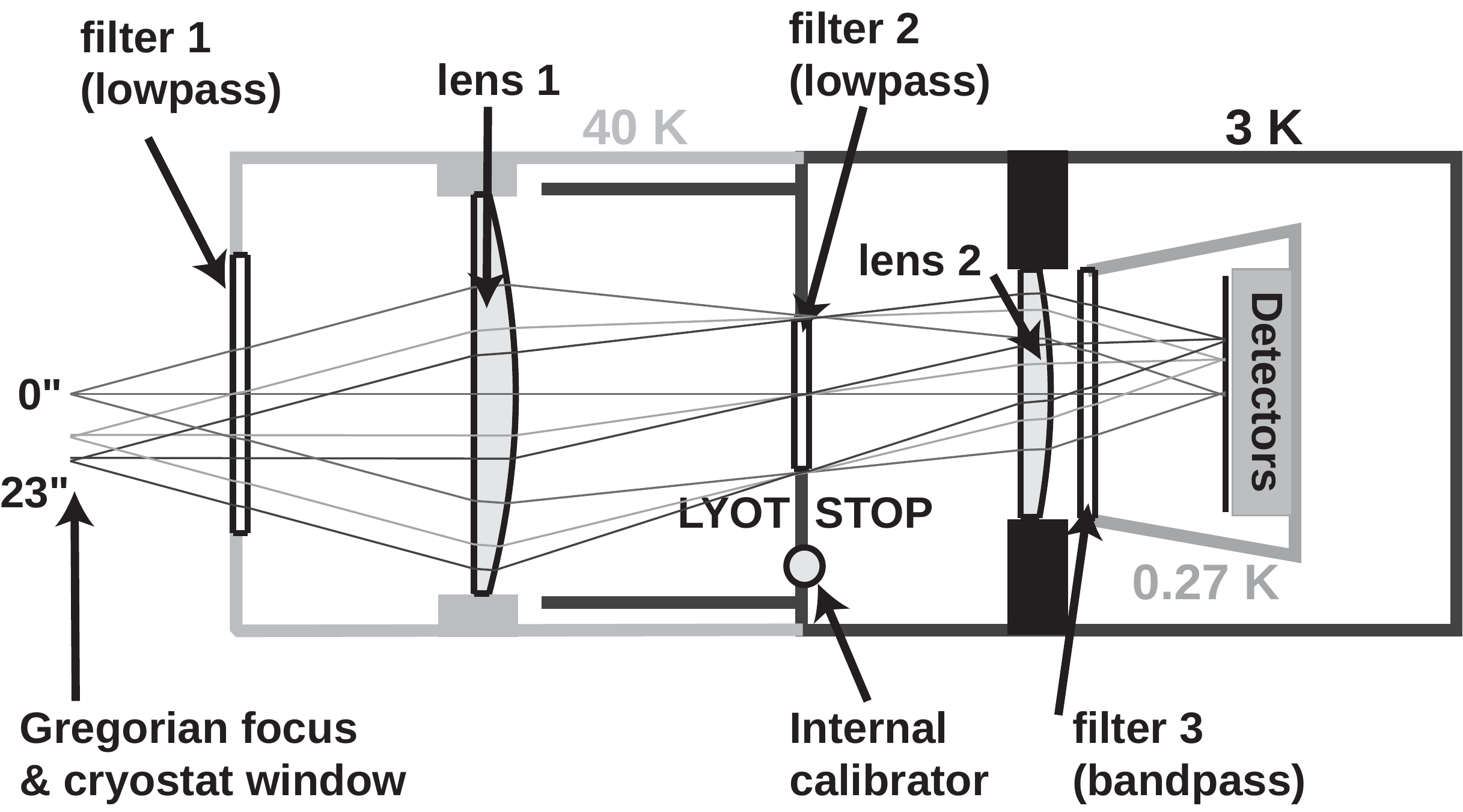}
   \end{tabular}
   \end{center}
   \caption[cryostat]{\label{fig:optical} MUSTANG's optical design.  HDPE lenses refocus the light from the Gregorian focus onto the array while capacitive mesh filters define the band.  The inside of all the housing for the lenses is painted black to absorb any stray light.}
\end{figure}

MUSTANG's optical design is shown in figure~\ref{fig:optical}. A
high density polyethylene (HDPE) lens forms an image of the primary
mirror of the GBT at the Lyot stop. The aperture of the Lyot stop
is chosen to
give uniform illumination of the GBT's primary out to a diameter of
90~m.  Calculations show that once diffraction is taken into account,
spillover at the primary is less than 1.6\%\cite{optics_paper}.  After
the Lyot stop a second lens reimages the Gregorian focus of the GBT at
$f/1.62$ so as to obtain a $4''$ $(0.5 f \lambda)$ pixel spacing on the
sky.  A small black body placed at the Lyot stop uniformly illuminates the
array.  During observations this source can be pulsed and used to
monitor changes in the response of detectors. To reduce optical loading on the
detectors, the first lens is cooled to 40~K and the second lens and the
Lyot stop are cooled to 3~K. 

MUSTANG's 81--99~GHz band-pass is defined by a capacitive mesh
filter placed over the array.  Other low-pass capacitive mesh filters
at the Lyot stop and in front of the first lens, block leaks in the
band-pass filter and, with the help of IR blocking filters, control the
optical loading on the cryogenics.  All these filters are made of
sheets of patterned 
metalized polypropylene at Cardiff University.  The band-pass
and low-pass filters are made of many layers hot-pressed together and are more robust
than traditional air-gap filters\cite{Ade2006}.

Care was taken to control stray light.  Stray light can cause a number
of 
problems. Excessive loading on detectors can cause them to saturate and
reflections between pairs of optical elements can reimage astronomical
sources onto different parts of the array (ghosting).  These problems
can be especially acute in on-axis optical designs.  To reduce
reflections, the lenses have $\lambda/4$ grooves in their surfaces.
Cold, light absorbing baffles are placed throughout the optical
system.  Simulations showed that no ghost image
with an amplitude greater than the diffraction limited point spread
function is formed on the array\cite{optics_paper}.

\section{INSTRUMENT PERFORMANCE}
Extensive lab characterization of the instrument has been carried out with
the detectors looking at 300~mK and with the detectors looking out
into the room.  After lab characterization  in late 2006, MUSTANG had its
first commissioning run on the GBT\@.  The results of this run were used
to make some improvements to the instrument, most significantly a
large reduction in the amount of 1/f noise.  In December 2007 we had
another period of commissioning followed, in the spring of 2008, by our
first science observations.  During the summer of 2008 upgrades to the
instrument are being carried out before next winter's observing season.

 \subsection{Lab Characterization}
Currently, MUSTANG has 57 pixels which show a good response to
 the internal calibrator a yield of 89\%.  Of the remaining pixels, five
 are electrically dead while two show a poor optical response and
 excessive noise.  The optical efficiency of
the receiver was measured using hot and cold loads and found to be 73\%.

Measurements of the transition temperatures of all working pixels gave
a mean $\rm T_c$ of 490~mk, 40~mK hotter than our target value.  There was
also a larger than expected scatter with some values over 500~mK and
others as low as 477~mK.  The dimensionless measure of the steepness
of the transitions, $\alpha$, was estimated to be 460, three to four
 times higher than expected. The saturation power of detectors was
also found to be higher than expected, typically 42~pW instead of
25~pW.

\begin{figure}
   \begin{center}
   \begin{tabular}{c}
   \includegraphics[height=11cm,angle=270]{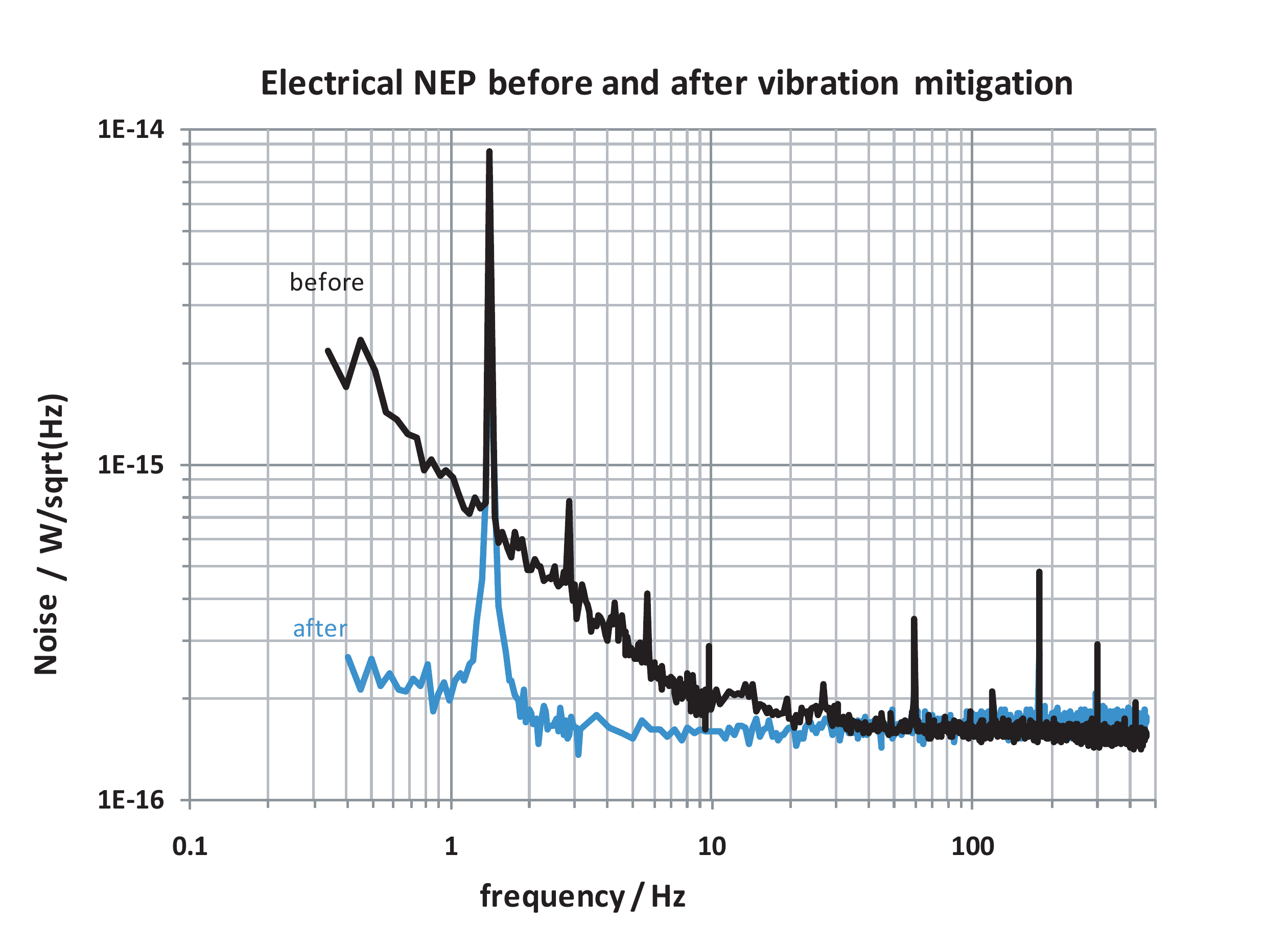}
   \end{tabular}
   \end{center}
   \caption[cryostat]{\label{fig:power} Examples of MUSTANG power
     spectra before and after the removal of the $1/f$ noise.  The
     noise spike at 1.4~Hz is due to temperature fluctuations of the
     pulse tube.  This spike is narrow and common mode in all pixels.}
\end{figure}

Noise power spectra measurements on good pixels gave electrical NEPs
between 20 and 110~Hz of around
$1.6\times10^{-16}\ \mbox{W}/\sqrt{\mbox{Hz}}$,
significantly above our target value.  In addition most pixels
had excess $1/f$ noise below 20~Hz. Below 20~Hz is where
 our astronomical signal is located so reducing the $1/f$ noise was a
priority.  The noise was shown to be 
uncorrelated between pixels.  It was eventually discovered that pixels
had narrow mechanical resonances at frequencies around 150~Hz with
different pixels having slightly different resonance peaks.  
Vibrational isolation of the pulse tube 
mitigated the problem. It is worth noting that  other experiments such
as ACT\cite{ACT} successfully use TES bolometers, cooled using pulse
tubes in similar cryostats to MUSTANG,
 without seeing this problem. It is likely that our
susceptibility to vibrations comes from the mechanical design of our
pixels. 
Examples of power spectra for some of our pixels can be seen in figure~\ref{fig:power}. 

A clear feature in the power spectra is a
narrow ($<$0.1~Hz) spike at 1.4~Hz. This is the frequency our pulse tube
operates at.  This spike is common mode to all pixels and
we attribute the majority of it to temperature fluctuations of the 3~K
optical components.  A smaller fraction of this noise 
can be traced to temperature
fluctuations of the series array SQUID amplifiers that are the last
stage of the NIST multiplexing electronics.  
As the 1.4~Hz noise is so narrow and
common mode it is easy to remove in data processing. In the
long term, passive damping of the temperature fluctuations at the head
of the pulse tube would be of benefit.  Active temperature control
would be less desirable as it requires extra power and has a greater
chance of failure. 

 In an effort to understand our higher than expected white noise
 between 20 and 110~Hz
 we measured the complex impedance and the response to bias 
 of our detectors. Simple models of our detectors that took into
 account thermal coupling between the absorber, the membrane, the TES,
 and the thermal bath were able to produce a qualitative fit to our data
 and predicted theoretical NEPs around $7 \times10^{-17}\
 \mbox{W}/\sqrt{\mbox{Hz}}$. Our model did not take into account
 factors such as a finite thermal conduction across our large
 membranes and possible hanging heat capacities, all of which can
 increase the noise.
  What was clear from our simple model
 was that many parameters such as the phonon conduction between the
 membrane and the bath differed from their target values in such a way
so as to increase the noise significantly above our target.

In summary, our lab tests showed that MUSTANG's current array is not
ideal, having a phonon noise level 8 times
higher than the background photon noise expected at Green Bank.
However, sensitivity calculations showed that MUSTANG had enough sensitivity
for useful preliminary science observations.  Now that these observations
 have been made we are looking into the design of a replacement array with better sensitivities.   

\subsection{On Telescope Performance}
Initial tests showed identical noise performance on the telescope as
in the lab.  The one exception was that at elevations less
than 30 degrees some of the $1/f$ noise returned.  In this case it
could be traced to the high vibration Gifford-McMahon refrigerators in other
receivers. Typically only a few detectors were affected at a time,
however, it is our intention to move to a different array design in
order to mitigate this problem. 

Measurements of bright point sources showed that MUSTANG had a round beam
shape of 8--10$''$\ FWHM, in good agreement with our optical models.  
By measuring the amplitude of Mars, the surface
efficiency for the GBT was found to be 12\%, consistent with the
surface RMS measured at Q-band (390~$\mu$m).
Deep beam maps show that most of the power lost due to the primary
mirror's RMS is being scattered to large angular scales or into a pedestal
around the beam.  This implies that
the most of the errors in the GBT surface are on small (panel to
panel) or medium
length scales not gross distortions of the dish that would distort
the main beam.  However out-of-focus
holography using MUSTANG was able to detect some large scale
distortions in the GBT's primary.  
Out-of-focus holography\cite{OOF} uses
data similar to what is currently taken when focusing MUSTANG\@. In the
future it should be possible to measure and remove all remaining large
scale distortions, typically produced by thermal effects, when
focusing MUSTANG\@.   Detector loading at $77^\circ$ was found to
vary between  4~pW in our best weather ($\tau=1$) and 9~pW when the
weather had a $\tau=0.2$.

\begin{figure}
   \begin{center}
   \begin{tabular}{c}
   \includegraphics[height=7cm]{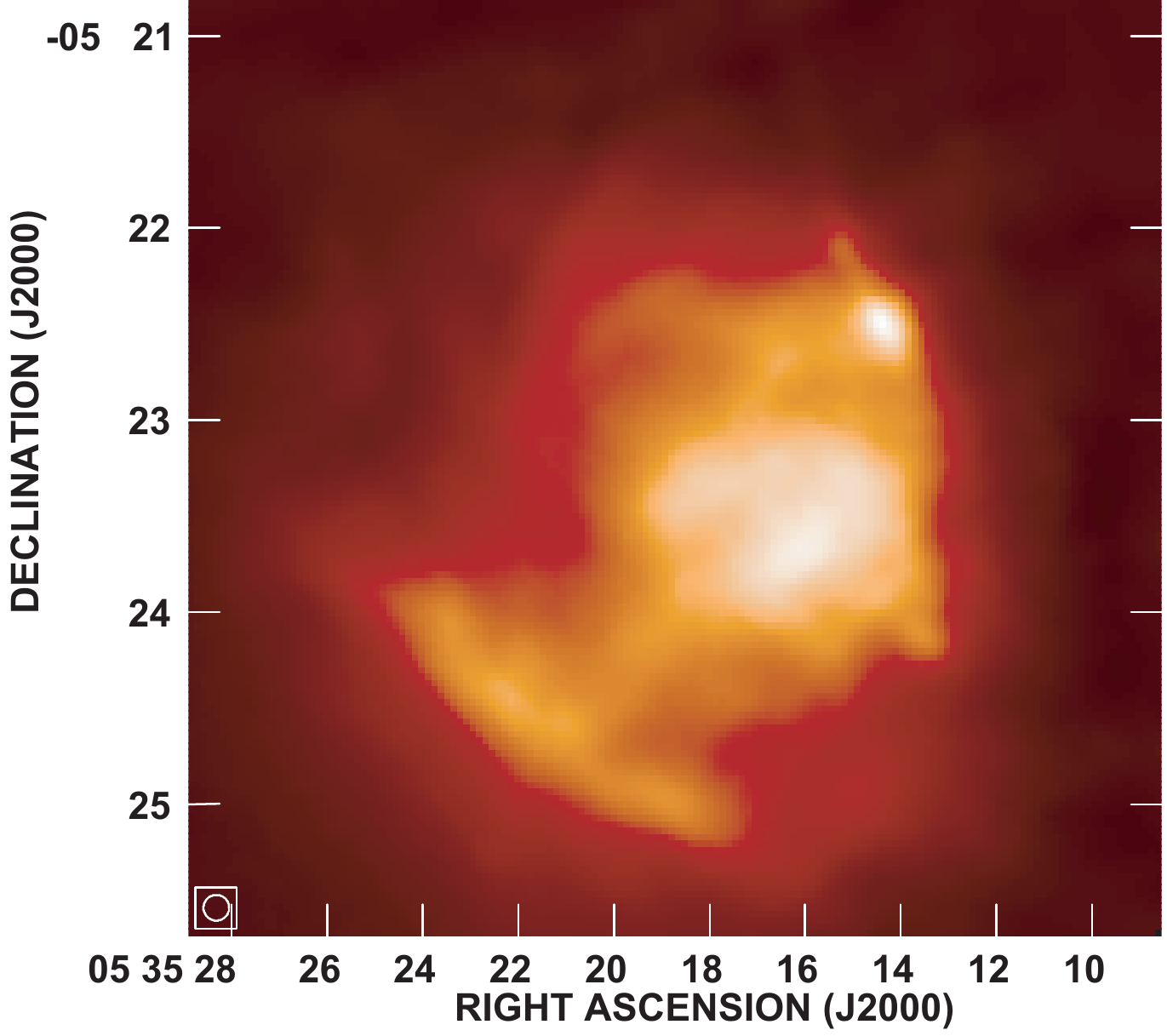}
   \includegraphics[height=7cm]{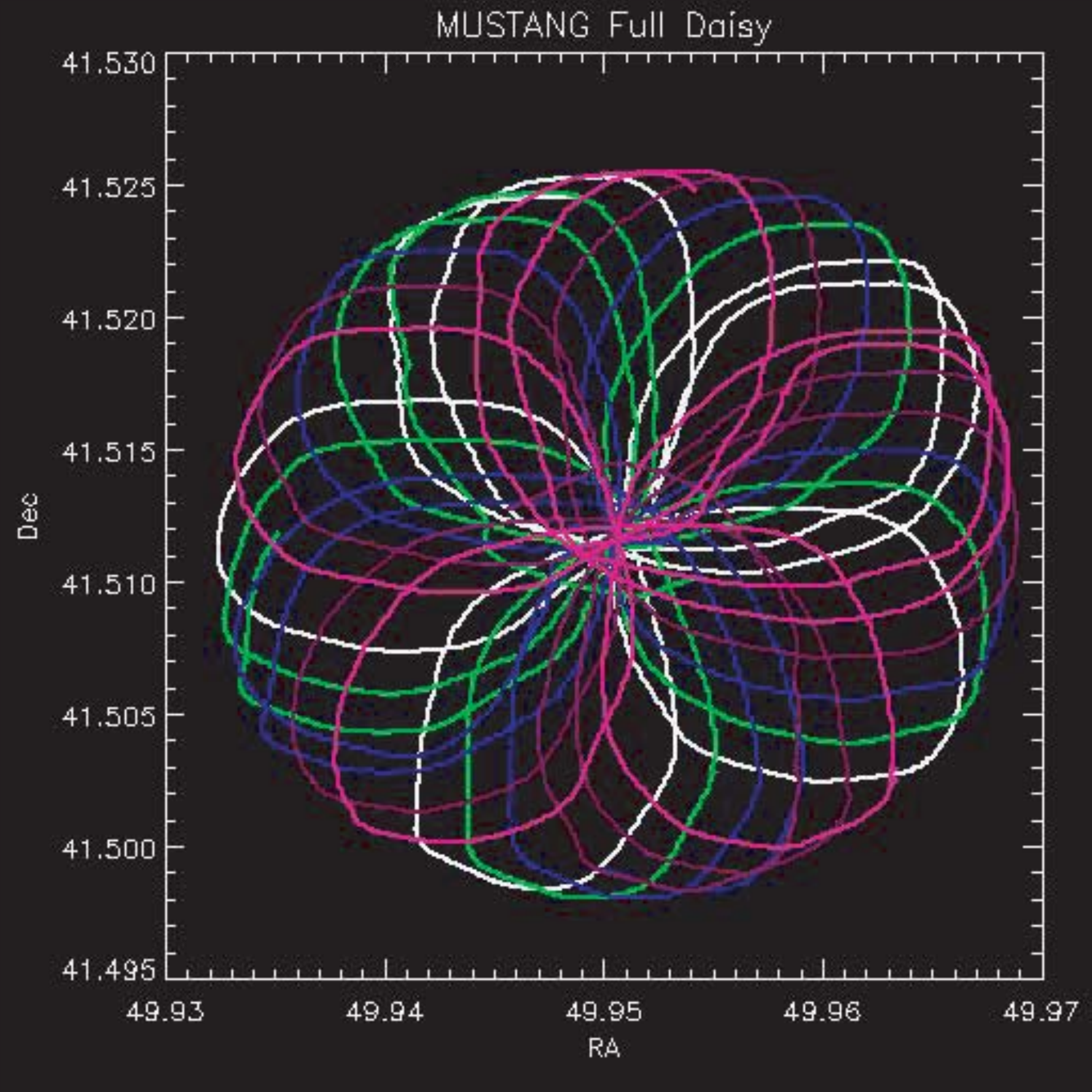}
   \end{tabular}
   \end{center}
   \caption[cryostat]{\label{fig:astronomy} Part of the star forming
     region around Orion KL. Shown in the bottom left
   of the picture is our beam size.  Observations were made
   using scanning patterns that produce lots of cross linking, an
   example of which is the daisy pattern shown on the right.  The
   daisy pattern has the advantage of low acceleration which minimizes the
   chance of exciting oscillations in the structure of the GBT.  It
   strongly concentrates coverage towards the center making it good
   for point sources.}
\end{figure}

When making total power measurements, any change in atmospheric emission
and any low frequency drifts in the receiver will produce large amounts
of noise in maps.  This is often removed by chopping between
positions on the sky.  With an array such as MUSTANG it is possible to
create scan patterns that make repeated observations of the same
part of the sky on many different time-scales. In a single pass, a point
source on the sky will only be in a few detectors at any moment in time
 so it is possible to
remove common mode noise such as our 1.4Hz signal which appears in all
 detectors simultaneously.  Multiple re-observations of
different patches sky on many different time scales (1--300 seconds)
 enable slow drifts
from the atmosphere or receiver to be removed.  We use two
such scan strategies, the daisy pattern shown in
figure~\ref{fig:astronomy} which is good for point source observations
and a basket weave pattern which gives more uniform coverage and is
good for mapping large areas.  An example of an observation made using
a basket weave pattern is shown in figure~\ref{fig:astronomy}.

It is worth noting that many aspects of MUSTANG and the GBT were
remarkably stable. The instrument was operated continuously on the GBT
for a period of over 3 months without incident. 
 The detector and SQUID biases could be reused with
minimal adjustment from night to night. Selection of optimal bias
values for the detectors and setting up the SQUID amplifiers
 typically takes 15 minutes.
 Gain changes, measured using
the black body source were less than 3.7\% during an eight hour
observing session.    

\section{CONCLUSIONS} 
MUSTANG has already demonstrated the potential of bolometer arrays on
large telescopes such as the GBT\@.  Over the coming year we will be
making a number of improvements to both the instrument and the GBT's
surface enabling more sensitive observations.  By achieving
a 240~$\mu$m RMS in the primary of GBT we could gain a factor of 3.5
in sensitivity.  Another
factor of 2 could be gained by replacing the current array with one
that has saturation powers closer to the maximum loading we expect.
  Expanding the array to 256 pixels
would only require changes to the detector package and the optical
elements and this is part of our long term plan. 
 NRAO have invited
others in the astronomical community to make proposals to use MUSTANG
on a shared risk basis.  In
order for it to fully become a functional facility instrument, work is
being carried out on the user interface.  In particular we hope to
incorporate the improved bias tuning software developed for the ACT
experiment\cite{tunning}. We will better integrate focusing routines
into the GBT's control system and develop the capability for rapid,
real-time, out-of-focus Holography which will correct for any large
scale aberrations of the primary.



\acknowledgments     

Funding for this project was provided by the National Radio Astronomy
Observatory through the NSF and NSF award number AST-0607654.

\bibliography{MUSTANG}   
\bibliographystyle{spiebib}   

\end{document}